\documentclass[aps,prl,showpacs,twocolumn,floats]{revtex4}
\usepackage{graphicx}
\usepackage{dcolumn}
\usepackage{bm}

\begin{document}

\title{Self-organized patterns of macroscopic quantum tunneling in molecular magnets}
\date{\today}
\author{D. A. Garanin and E. M. Chudnovsky}
\affiliation{Physics Department, Lehman College, City University
of New York \\ 250 Bedford Park Boulevard West, Bronx, New York
10468-1589, USA}
\date{5 December 2008}

\begin{abstract}
We study low temperature resonant spin tunneling in molecular magnets
induced by a field sweep with account of dipole-dipole interactions.
Numerical simulations uncovered formation of self-organized patterns of the
magnetization and of the ensuing dipolar field that provide resonant condition inside a
finite volume of the crystal. This effect is robust with respect to disorder
and should be relevant to the dynamics of the magnetization steps observed
in molecular magnets.
\end{abstract}
\maketitle

\begin{figure}[tbp]
\vspace{-0.5cm} \includegraphics[width=5.5cm,angle=-90]{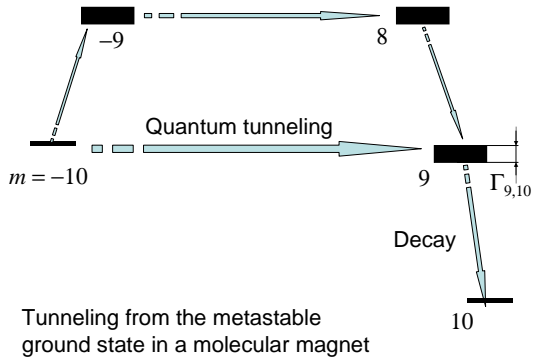}
\caption{Thermally assisted tunneling.}
\label{levels}
\end{figure}

Molecular magnets had become focus of interest after it was demonstrated
\cite{sesgatcannov93nat} that a crystal of Mn$_{12}$-Acetate molecules is
equivalent to a regular array of single-domain superparamagnetic particles
of spin 10. The subsequent discovery of the stepwise hysteresis curve in Mn$%
_{12}$-Ac \cite{frisartejzio96prl} made it a model system for the
study of macroscopic quantum tunneling of the magnetic moment
\cite{chutej98book}. In a typical macroscopic experiment the
magnetic field is swept at a constant rate, making the spin energy
levels $|m\rangle$ and $|m^{\prime}\rangle$ cross at certain
fields $B = B_k$, see Fig.\ \ref{levels}.
Probability of a quantum transition between the crossing levels affects the
size of the magnetization step. In this Letter we develop theoretical
framework for understanding the dynamics of the steps. We will be concerned
with the regime of pure quantum tunneling or thermally assisted quantum
tunneling \cite{garchu97prb}.

It had been recognized in the past \cite{prosta98prl} that the
knowledge of the probability of spin tunneling alone is
insufficient to describe the tunneling dynamics of molecular
magnets. This is because the reversal of any individual spin
changes the long-range dipolar magnetic field of the entire
crystal, which tunes other spins in or out of resonance. The
tunneling dynamics of a molecular magnet near a resonance is,
therefore, non-local and non-linear \cite{gar-dip}. Prokof'ev and
Stamp \cite{prosta98prl} suggested that the relaxation could occur
near some surface inside the sample which is the locus of points
characterized by the zero bias between spin energy levels. Their
local model, however, could not capture the evolution of such a
surface in time that should occur during magnetic relaxation \cite
{chu00prl-comment}. Later studies of the collective relaxation in
molecular magnets \cite{cucetal99epjb,Fernandez-Stamp} did take
the non-locality into account. They did not consider, however, the
dynamics generated by the field sweep that is important for the
formation of a coherent spatial structure in the macroscopic
tunneling dynamics driven by dipolar fields.

In this Letter we derive and solve non-linear integro-differential
equations that govern the self-organized tunneling dynamics of
spins coupled by dipolar interactions.

Let $n_{m}$ be the normalized population of the $m$-th level, $\sum_m n_m =1$%
. In a typical macroscopic experiment one magnetizes the crystal
such that initially $n_{-S}=1$ while populations of other states
are zero. The field is then swept in the opposite direction at a
constant rate. It has been demonstrated theoretically
\cite{Quant-Stat-Theory} and confirmed by experiments
\cite{Quant-Stat-Exp} that in the thermally assisted regime a step
in the magnetization curve is dominated by a resonance between
particular $m$ and $m^{\prime}$ levels, see Fig.\ (\ref{levels}).
This
resonance is determined by the maximum of $\Delta^2_{mm^{\prime}}\exp[%
-(E_m-E_{-S})/(k_BT)]$, where $\Delta_{mm^{\prime}}$ is the tunnel splitting
of the $m$-th and $m^{\prime}$-th levels, $E_m$ is the energy of the $m$-th
level, and $T$ is temperature. On approaching this resonance, the population
of the $|-S\rangle$ state satisfies the following equation \cite{garchu97prb}
\begin{equation}
\dot{n}_{-S}=-\Gamma n_{-S}  \label{dotrhoEq}
\end{equation}
where
\begin{eqnarray}
\Gamma & = & \frac{1}{2}\left(\frac{\Delta_{mm^{\prime}}}{\hbar}%
\right)^2\exp\left(-\frac{E_m-E_{-S}}{k_BT}\right)  \nonumber \\
& \times & \frac{\Gamma _{m,m^{\prime }}/2}{\left( W/\hbar \right)
^{2}+\left( \Gamma _{m,m^{\prime }}/2\right) ^{2}}\,.  \label{GammaDef}
\end{eqnarray}
Here $W$ is the energy bias between $m$ and $m'$, and $\Gamma
_{m,m^{\prime }} \equiv \Gamma _{m} + \Gamma _{m^{\prime}}$,
$\Gamma _{m}$ being the rate of transitions from $m$ to the
neighboring spin levels. The overdamped condition $\Gamma
_{m,m^{\prime }} \gg \Delta$ is assumed, which is the case of
practical interest. Since the $|m^{\prime}\rangle$ state quickly
decays down to the $|S\rangle$ state, the magnetization is given
by
\begin{equation}
\sigma_z \equiv \langle S_z \rangle/S = 1-2n_{-S}\,.
\end{equation}
At a constant sweep rate, $W = v_W t$, when $\Gamma_{m,m^{\prime}}$ is small
compared to the distance between adjacent spin levels, one can integrate
Eq.\ (\ref{dotrhoEq}) from $-\infty$ to $\infty$ to obtain the population of
the $|-S\rangle$ state after the level crossing. This gives the Landau-Zener
like asymptotic result,
\begin{equation}  \label{LZ}
n_{-S}(\infty) = e^{-\varepsilon}\,,
\end{equation}
with $\varepsilon$ corrected for thermally assisted tunneling \cite
{garchu02prb},
\begin{equation}  \label{LZ-epsilon}
\varepsilon \equiv \frac{\pi \Delta^2_{mm^{\prime}}}{2\hbar v_W}\exp\left(-%
\frac{E_m-E_{-S}}{k_BT}\right)\,.
\end{equation}
The first magnetization step occurs when the tunnel splitting of
the two states is sufficient to provide a not very small
$\varepsilon$ for a given sweep rate $v_W$. All of the above
formulas apply to the case of pure quantum tunneling that occurs
at $T = 0$. In the latter case $m = -S$.

The above simple picture contradicts experiments. Indeed, according to Eq.\ (%
\ref{GammaDef}) the width of the magnetization step due to the relaxation
dynamics described by Eq.\ (\ref{dotrhoEq}) is determined by $\Gamma
_{m^{\prime}}$. Such a width is very small compared to the experimental
value. One could naively think that the relaxation of any individual
molecule is still determined by Eqs. \ (\ref{dotrhoEq}) and (\ref{GammaDef})
but the resonant field is spread due to inhomogeneous dipolar and hyperfine
fields, and due to structural disorder. This, however, cannot explain the
height of the magnetization step as this would typically result in the total
magnetization reversal during the first step.
As we shall see, the non-trivial space-time dependence of the local energy
bias $W$ due the evolution of the local dipolar field $\mathbf{B}^{(\mathrm{D%
})}$ solves this controversy.

To simplify the problem we stick to the geometry of a long cylinder of
length $L$ and radius $R \ll L$, and restrict our consideration by the
longitudinal components of the fields. The transverse component of the field
enters the problem through its known effect on $\Delta_{mm^{\prime}}$. The
bias at the cite $i$ is given by
\begin{equation}
W_{i}=\left(m^{\prime }-m\right) g\mu _{\mathrm{B}}\left( B-B_{k}+B_{i}^{(%
\mathrm{D})}\right) \equiv W_{\mathrm{ext}}+W_{i}^{(\mathrm{D})},
\label{WiDef}
\end{equation}
where $B_{k}$ is the resonant value of the external field for the $k$-th
tunneling resonance ($k=S-m^{\prime }$) in the absence of the dipolar field.
The dipolar component of the bias is given by
\begin{equation}
W_{i}^{(\mathrm{D})}=\left( 1-\frac{m^{\prime }}{m}\right) E_{\mathrm{D}%
}D_{i},\qquad D_{i}\equiv \sum_{j}\phi _{ij}\sigma_z^j,  \label{WiDDef}
\end{equation}
where $E_{\mathrm{D}}\equiv \left( g\mu _{\mathrm{B}}S\right) ^{2}/v_{0}$ is
the dipolar energy, $E_{\mathrm{D}}/k_{\mathrm{B}}=0.0671$ K for Mn$_{12}$%
-Ac, $v_{0}$ is the unit-cell volume, and
\begin{equation}
\phi _{ij}=v_{0}\frac{3\left( \mathbf{e}_{z}\cdot \mathbf{n}_{ij}\right)
^{2}-1}{r_{ij}^{3}},\qquad \mathbf{n}_{ij}\equiv \frac{\mathbf{r}_{ij}}{%
r_{ij}}  \label{phizijDef}
\end{equation}
is the dimensionless dipole-dipole interaction between the spins at the
sites $i$ and $j\neq i.$

\begin{figure}[tbp]
\includegraphics[angle=-90,width=9cm]{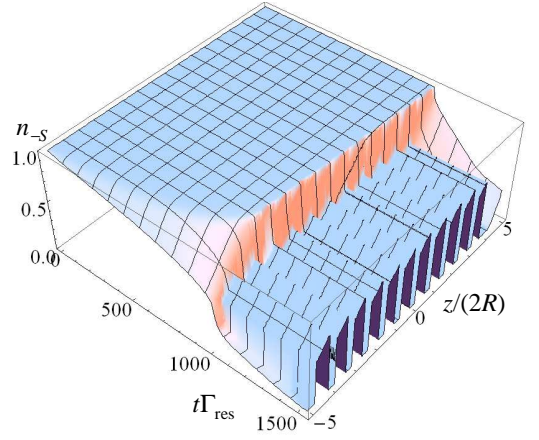}
\caption{Color online: Moving wall of resonant spin tunneling induced by a
slow sweep, $\protect\varepsilon =10.$ There are well-developed spatially
quasiperiodic, time independent structures behind the wall.}
\label{Fig-n_3d_Wext-sweep}
\end{figure}
To write down the dynamical equation for the entire sample, it is convenient
to use dimensionless variables
\begin{equation}
\widetilde{t}\equiv \Gamma _{\mathrm{res}}t\,,\quad \widetilde{z}=\frac{z}{R}%
\,,\quad \widetilde{W}\equiv \frac{W}{(1-m^{\prime }/m)E_{\mathrm{D}}}\,,
\label{WtilDef}
\end{equation}
where
\begin{equation}
\Gamma _{\mathrm{res}}=\frac{\Delta _{mm^{\prime }}^{2}}{\hbar ^{2}\Gamma
_{m,m^{\prime }}}e^{-(E_{m}-E_{-S})/(k_{B}T)}\,,  \label{GammaResDef}
\end{equation}
is the relaxation rate on resonance, see Eq.\ (\ref{GammaDef}).
Equation (\ref{dotrhoEq}) becomes a non-linear
integro-differential equation,
\begin{equation}
\frac{d}{d\widetilde{t}}\,n_{-S}(\widetilde{z},\widetilde{t})=-F(\widetilde{z%
},\widetilde{t})n_{-S}(\widetilde{z},\widetilde{t})\,.  \label{dnduEq}
\end{equation}
Here $F$ contains integral dependence on $n_{-S}(\widetilde{z},\widetilde{t})
$ via $\widetilde{W}$,
\begin{equation}
F(\widetilde{z},\widetilde{t})=\frac{1}{1+4\widetilde{E}_{\mathrm{D}}^{2}%
\widetilde{W}^{2}(\widetilde{z},\widetilde{t})}\,,\qquad \widetilde{E}_{%
\mathrm{D}}\equiv \frac{(1-m^{\prime }/m)E_{\mathrm{D}}}{\hbar \Gamma
_{m,m^{\prime }}}\,.  \label{FDef}
\end{equation}
Summation of contributions from all sites in Eq.\ (\ref{WiDDef}) gives the
following expression for a continuous variable $\widetilde{W}(\widetilde{z},%
\widetilde{t})$:
\begin{eqnarray}
\widetilde{W} &=&\widetilde{W}_{\mathrm{ext}}+\widetilde{W}^{(D)}=\widetilde{%
W}_{\mathrm{ext}}-\kappa \lbrack 1-2n_{-S}(\widetilde{z},\widetilde{t})]
\nonumber \\
&-&2\pi \nu \int_{-L/(2R)}^{L/(2R)}d\widetilde{z}^{\prime }\frac{1-2n_{-S}(%
\widetilde{z}^{\prime },\widetilde{t})}{\left[ \left( \widetilde{z}^{\prime
}-\widetilde{z}\right) ^{2}+1\right] ^{3/2}}\,,  \label{WzzCylinder}
\end{eqnarray}
where $\nu $ is the number of molecules per unit cell and
\begin{equation}
\kappa =\frac{8\pi \nu }{3}-\sum_{j}\phi _{ij}\,.
\end{equation}
The summation in the last formula goes over the volume of a large sphere.
For the body-centered tetragonal lattice of Mn$_{12}$-Ac one has $\nu =2$
and $k=14.6$ \cite{garchu08prb}. In the depth of a uniformly polarized
elongated sample Eq.\ (\ref{WzzCylinder}) provides the dipolar field of $526$%
G in good agreement with the measured value of $515\pm 85$G \cite{Sean}.
Equations (\ref{dnduEq})-(\ref{WzzCylinder}) should be solved with the
initial condition $n_{S}(t=-\infty )=1$.

For Mn$_{12}$-Ac $\widetilde{E}_{\mathrm{D}}$ is a large parameter, $%
\widetilde{E}_{\mathrm{D}} > 100$, so that one can naively think that $F$ is
always small except for a very brief period of time when the field sweep
brings the bulk of the sample close to the resonance. If this was true, the
total relaxation would have been vanishingly small. However, as we shall see
below, the system finds the way to relax faster by forming a moving wall of
finite width $l \sim R$ inside which $\widetilde{W}$ is so small that $F$ is
not significantly reduced from its maximal value $F = 1$. Inside the wall
region the resonant tunneling transitions take place. The greater is $%
\widetilde{E}_{\mathrm{D}}$, the closer to zero is the energy bias $%
\widetilde{W}$ in the wall. Beyond the wall region $\widetilde{W}$ deviates
strongly from zero, $F$ becomes very small, and the relaxation effectively
freezes.

\begin{figure}[tbp]
\includegraphics[angle=-90,width=8cm]{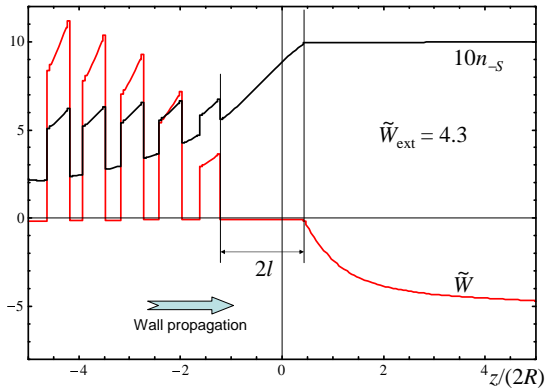}
\caption{Color online: Spatial profiles of the metastable population $n_{-S}$
and the reduced bias $\widetilde{W}$ at $\widetilde{W}_{\mathrm{ext}}=4.3$
and $\widetilde{E}_{\mathrm{D}}=200.$ Everywhere in the wall the system is
near the resonance, $\widetilde{W}\approx 0.$ }
\label{Fig-n_W_profiles_Wext=4.3}
\end{figure}

\begin{figure}[tbp]
\includegraphics[angle=-90,width=8cm]{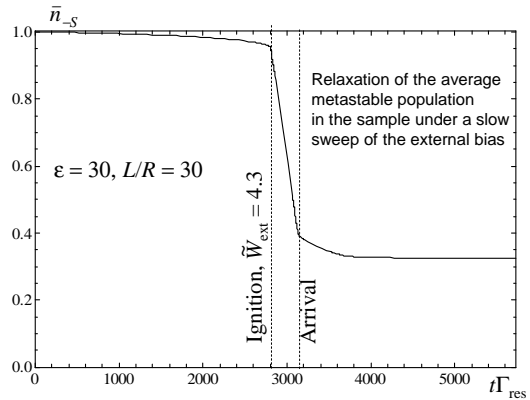}
\caption{{}Time dependence of the average metastable population $\bar
{n}_{-S} = \langle n_{-S}\rangle$.}
\label{Fig-n_avr_Wext-sweep}
\end{figure}

Numerical solution of Eq.\ (\ref{dnduEq}) based upon its
discretization over the length of the sample yields a propagating
wall of tunneling shown in Fig.\ \ref{Fig-n_3d_Wext-sweep}.
Spatial dependence of the metastable population $n_{-S}$ and of
the energy bias are shown in Fig.\ \ref
{Fig-n_W_profiles_Wext=4.3}. Striking universal features of the
relaxation process have been uncovered by the simulations. The
wall ignites as the bias reaches the ``magic'' value of
$\widetilde{W}_{\mathrm{ext}}=4.3$, which corresponds to $B = B_k
+ 194$G in Mn$_{12}$-Ac regardless of the sweep rate and the level
$m$ that dominates the transition. One can see from Fig.\ \ref
{Fig-n_W_profiles_Wext=4.3} that everywhere in the wall the system
is near the resonance, $W \approx 0$.
The reduced speed of the wall,
\begin{equation}
v^{\ast }\equiv \frac{v}{R\Gamma _{\mathrm{res}}}\,,  \label{vstarDef}
\end{equation}
depends on the parameter $\widetilde{E}_{\mathrm{D}}$. The propagation of
the wall does not result in the total magnetization reversal. The metastable
population behind the wall initiated by a slow sweep is a universal number $%
n_{f}\approx 0.32$, see Fig.\ \ref{Fig-n_avr_Wext-sweep}.
The width of the wall, $2l$, is of the order of the diameter of the
elongated sample. Figs.\ \ref{Fig-n_3d_Wext-sweep} and \ref
{Fig-n_W_profiles_Wext=4.3} show the tunneling wall and the quasiperiodic
structure behind the wall in the left half of the sample. In a totally
symmetric sample, two walls will simultaneously ignite on the left and on
the right sides of the sample, and then move towards the center. Deviation
from the dynamics depicted in the above figures will occur only when the
distance between the two walls becomes comparable to $R$. For $L \gg R$ this
should not change our conclusions.

\begin{figure}[tbp]
\includegraphics[angle=-90,width=8cm]{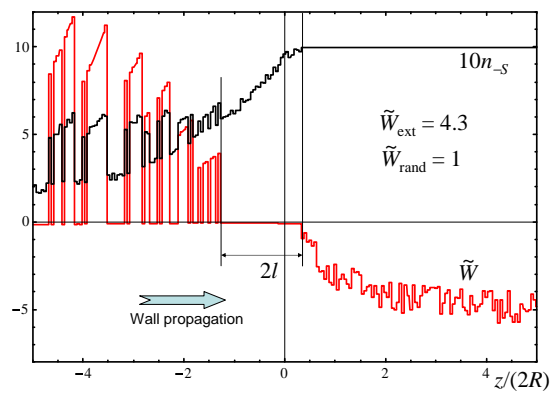}
\caption{Color online: Spatial profiles of the metastable population $n_{-S}$
and the reduced bias $\widetilde{W}$ in the presence of a relatively strong
disorder. The condition $\widetilde{W} \approx 0$ inside the wall is not
affected by disorder.}
\label{Fig-n_W_profiles_Wext=4.3_Wrand=1}
\end{figure}

So far we have not considered the effect of disorder on the
formation of the tunneling wall. For Mn$_{12}$-Ac this question is
important because nuclear spins, solvent disorder, and crystal
defects produce randomness in the local values of $B_k$ and local
dipolar fields. To study the effect of disorder we added a random
component, $W_{\mathrm{rand}}$, to the energy bias at each lattice
site. While disorder produces visible fluctuations of $n_{-S}$ and
$W$ behind the wall and ahead of the wall, the existence of the
wall with $W \approx 0$ is not affected by disorder, see Fig.\
\ref{Fig-n_W_profiles_Wext=4.3_Wrand=1}.

We believe that our results are relevant to the magnetization
steps commonly observed in molecular magnets at a slow field
sweep. These results are robust with respect to other activation
scenarios, e.g. those involving manifolds with a different total
spin $S$ \cite{Carretta}. The latter will simply modify the
expression for $\Gamma_{\rm res}$, Eq.\ (\ref{GammaResDef}).

Without dipolar fields the energy width of the magnetization step, according
to Eq.\ (\ref{GammaDef}), is $\Delta W_{\Gamma}=\hbar \Gamma
_{m,m^{\prime}}/2$. With account of the dipolar fields it is determined by
the time it takes the tunneling wall to cross the sample, $t = L/v$. During
that time the energy bias changes by $\Delta W = v_W t= v_W L/v =
v_WL/(v^{\ast}R\Gamma _{\mathrm{res}})$. This gives the following ratio of
the field widths of the magnetization step with and without the dipolar
fields:
\begin{equation}  \label{ratio}
\frac{{\Delta B}}{\Delta B_{\Gamma}} = \frac{\pi}{v^{\ast}\varepsilon}\left(%
\frac{L}{R}\right)\,,
\end{equation}
where we have used Eqs.\ (\ref{LZ-epsilon}) and (\ref{GammaResDef}). For the
parameters of Mn$_{12}$-Ac, in the limit of a very slow sweep, one
numerically obtains $v^{\ast} \sim 10^{-3}$. At $\varepsilon = 30$ and $L/R
= 30$ used in the simulations, Eq.\ (\ref{ratio}) then gives ${\Delta B}/{%
\Delta B_{\Gamma}} \sim 10^3$. This brings the width of the step
in the range that is usually observed in experiment. So far we
have studied the dynamics of the first magnetization step. The
second step begins with a quasiperiodic structure of magnetization
left after the first step. This dynamics will be studied elsewhere
based upon the equations derived above.

After this work was completed, it had come to our attention that
the oscillating magnetization near tunneling resonance, depicted
in Fig.\ \ref{Fig-n_W_profiles_Wext=4.3}, has been already seen in
experiment \cite{Avraham}. We encourage experimentalists to
perform detailed local measurements of molecular magnets to
confirm that the magnetization step involves propagating walls of
quantum spin tunneling.

This work has been supported by the NSF Grant No. DMR-0703639.

\end{document}